\documentstyle[aps,pre,floats,epsf]{revtex}
\begin{document}
\advance\textheight by 2mm
\draft

\twocolumn[\hsize\textwidth\columnwidth\hsize\csname@twocolumnfalse%
\endcsname


\title{ Numerical Study of Local and Global Persistence 
        in Directed Percolation} 

\author{Haye Hinrichsen$^1$ and Hari M. Koduvely$^2$}

\address{$^1$ Max-Planck-Institut f\"ur Physik komplexer Systeme,
         N\"othnitzer Stra\ss e 38, D-01187 Dresden, Germany\\
         $^2$ Department of Physics of Complex Systems, 
         Weizmann Institute of Science, Rehovot 76100, Israel} 

\date{November 25, 1997}

\maketitle

\begin{abstract}
The local persistence probability $P_l(t)$ that a site never becomes
active up to time $t$, and the global persistence probability $P_g(t)$
that the deviation of the global density from its mean value
$\rho(t)-\langle \rho(t) \rangle$ does not change its sign up to
time~$t$ are studied in a one-dimensional directed percolation process
by Monte Carlo simulations. At criticality, starting from random
initial conditions, both $P_l(t)$ and $P_g(t)$ decay algebraically
with exponents $\theta_l \approx \theta_g \approx 1.50(2)$, which is
in contrast to previously known cases where $\theta_g <
\theta_l$. The exponents are found to be independent of the initial
density and the microscopic details of the dynamics, suggesting that
$\theta_l$ and $\theta_g$ are universal exponents. It is shown that in
the special case of directed-bond percolation, $P_l(t)$ can be related
to a certain return probability of a directed percolation process with
an active source (wet wall).
\end{abstract}

\pacs{PACS numbers: 64.60.Ak, 05.40.+j, 05.70.Ln}
]  


\def\transfer{\mbox{\bf T}}
\def\smat{\mbox{\bf S}}
\def\pmat{\mbox{\bf P}}
\def\wmat{\mbox{\bf W}}
\def\umat{\mbox{\bf U}}

\section{Introduction}
\label{IntroSection}

In recent years it has been realized that certain first passage
quantities in nonequilibrium systems exhibit a power law decay with
non-trivial exponents
\cite{DBG,BDG,Cardy,DHP,MSBC,DHZ,MBCS,Krug,SZ,NMGO}.  One of these
quantities is the local persistence probability $P_l(t)$, defined as the
probability that a local variable at a given space point (normally a
spin) has not changed its state until time~$t$ during a stochastic
evolution. In various systems it was found that $P_l(t) \sim
t^{-\theta_l}$, where $\theta_l$ is called {\em local persistence
exponent}. A similar quantity, the global persistence probability
$P_g(t)$, defined as the probability that the global order parameter
does not change its sign up to time $t$ is also found to decay as a
power law with a {\em global persistence exponent} $\theta_g$. In
general the exponents $\theta_l$ and $\theta_g$ are found to be
independent of the usual scaling exponents and different from
each other. The persistence probabilities depend on the history of 
evolution as a whole, and thus it is generally hard to determine
these exponents analytically. Only a few cases of exact results are
known~\cite{BDG,DHP}.

An important nonequilibrium process which has been studied 
extensively is directed percolation (DP)~\cite{DPReview}. 
A large variety of physical systems which undergo a
phase transition from a fluctuating active phase into an absorbing
state (i.e. a configuration once reached, the system cannot escape
from) belongs to the same universality class as DP. Therefore it would 
be interesting to study the persistence probability in DP since
this could provide some understanding of the non-equilibrium nature of
this process.

In this article we present a numerical study of the local and the
global persistence probabilities in one-dimensional ($1$-d) DP
processes.  Our results indicate that both quantities decay at the
critical point as a power law with exponents $\theta_l \approx
\theta_g \approx 1.5$. However, at present it is not clear
whether these exponents are independent 
of the known scaling exponents of DP. That 
$\theta_l$ and $\theta_g$ are the same within numerical errors
is surprising since in all previously known cases it was found that
$\theta_g < \theta_l$~\cite{MBCS}.

Very little is known about universal properties of persistence
exponents. In a DP process starting from random initial conditions we
find the value of $\theta_l$ to be independent of the initial
density, which is in contrast to the case of the $1$-d Glauber model
where $\theta_l$ depends explicitly on the initial
magnetization~\cite{DHP}. In addition we find these exponents to be
the same for different microscopic realizations of the DP process,
indicating that $\theta_l$ and $\theta_g$ are indeed universal
exponents.

We also show that the local persistence probability is related to
certain observables measured in $1$-d DP processes with boundaries.
Introducing a transfer matrix formalism we establish an exact relation
between $P_l(t)$ and a specific quantity measured in a DP process with
an absorbing boundary (dry wall). A similar relation is found between
$P_l(t)$ (which can be seen as a `first return'
probability) and the probability $R(t)$ that a DP process with a
steady active source (wet wall) returns to its initial conditions.
For a particular realization of DP, called bond-directed 
percolation, this relation can be proven exactly.
Thus the problem of local persistence is related
to both the dry and the wet wall problems in DP which have been
discussed in Refs.~\cite{EGJT96,LSMJ97,FHL97}.

The article is organized as follows: In Sect.~\ref{Overview} we briefly
review the DP process. Our numerical results are presented in
Sect.~\ref{NumericalSection}. Performing Monte Carlo (MC) simulations
we measure local and global persistence probabilities and estimate the
corresponding exponents $\theta_l$ and $\theta_g$.  In order to
measure $\theta_g$ more accurately we use a block spin method which
has been introduced recently in Ref.~\cite{BlockSpinMethod}.  In
Sect.~\ref{AbsorbingBoundarySection} the relation between $P_l(t)$ 
and a specific observable in a DP process with a {\it
dry} wall is proved exactly. 
The relation between $P_l(t)$ and the return probability
$R(t)$ in presence of an active source is discussed in
Sect.~\ref{ActiveSourceSection}.

\section{Directed percolation -- a brief overview}
\label{Overview}
Directed percolation~\cite{DPReview} is used as a model for the
spreading of some, generally non-conserved agent and plays a
role for certain autocatalytic chemical reactions and
the spreading of epidemics. In DP models, sites of a lattice are
either occupied by a particle (active, wet) or empty (inactive,
dry). In the dynamic processes particles can either self-destruct or
produce an offspring at a neighboring empty site. If the rate for
offspring production $p$ is very low, the system will always reach a
state without particles which is the absorbing state of the system. On
the other hand, when $p$ exceeds a critical value $p_c$, another
steady state of the system exists on the infinite lattice, 
where the density $\rho(p)$ of active sites is finite. Between the two
phases a continuous phase transition takes place which is
characterized by long range correlations. There are various different
models for DP, e.g.  directed site and bond percolation on a
lattice~\cite{DPReview}, cellular automata as the Domany-Kinzel
model~\cite{DomanyKinzel}, and the contact
process~\cite{ContactProcess}, to name only a few.

One of the most important properties of directed percolation is its
robustness with respect to the microscopic dynamics of the system.
According to a widely accepted conjecture formulated by Janssen and
Grassberger \cite{DPConjecture}, any transition from a fluctuating
active phase into a single, non-fluctuating and non-degenerate
absorbing state belongs to the DP universality class, provided 
the dynamical processes are local and characterized by a one-component
order parameter without special attributes like additional
symmetries or frozen randomness.  The DP universality class is
characterized by three critical exponents, namely the density exponent
$\beta$ and the scaling exponents $\nu_{||}$ and
$\nu_\perp$. Therefore, if $P_l(t)$ and $P_g(t)$ actually decay as a
power law, an interesting question would be whether $\theta_l$ and
$\theta_g$ are independent of these three exponents and exhibit a
similar robustness.

In the present work we analyze the persistence probabilities $P_l(t)$
and $P_g(t)$ by Monte Carlo simulations.  A DP model which is
convenient for this purpose is the Domany-Kinzel (DK) cellular
automaton \cite{DomanyKinzel}. The DK model is defined as follows: a
binary variable $\sigma_i(t)=0,1$ characterizes the state of site $i$
at discrete time~$t$. $\sigma=1$ means that the site is active (wet)
whereas $\sigma=0$ means that it is inactive (dry).  The automaton
evolves by a parallel update rule in which the state of
$\sigma_i(t+1)$ is selected according to transition probabilities
$\tau(\sigma_i(t+1) | \sigma_{i-1}(t),\sigma_{i+1}(t))$.  The
transition probabilities of the DK model are
\begin{eqnarray}
\label{TransitionProbabilities}
\tau(1|0,0)&=&0\nonumber\\
\tau(1|0,1)=\tau(1|1,0)&=&p_1\\
\tau(1|1,1)&=&p_2\nonumber\\
\tau(0|\sigma_{i-1},\sigma_{i+1})&=&
1-\tau(1|\sigma_{i-1},\sigma_{i+1}).\nonumber
\end{eqnarray}
Thus the model is controlled by two parameters $p_1$ and~$p_2$.  For
fixed $p_2<1$ there is a critical value $p_{1,c}$ where a continuous
phase transition takes place. It is assumed that for $p_2<1$ the
critical behavior along the whole line $p_{1,c}(p_2)$ is the same as
that of DP. In particular, the phase transition line includes
three particular points, namely~\cite{DKTransitionPoints}:
\begin{enumerate}
\item directed bond percolation: 
      $p_1=0.644701(1)$, \\ $p_2=2p_1-p_1^2$
\item directed site percolation: $p_1=p_2=0.705485(5)$
\item Wolfram's rule 18: $p_1=0.8092(1)$, $p_2=0$ 
\end{enumerate}
As a necessary condition, a universal critical exponent should not
depend on the choice of the transition point used in a simulation. 

Another special case is {\em compact} directed percolation where
$p_1=1/2$ and $p_2=1$. Here the dynamics of the model belong to
a different universality class, namely that of annihilating
random walks or, equivalently, the Glauber Ising model. In fact, it
turns 
out that in the case of compact DP the results of 
Refs.~\cite{DBG,DHP,MBCS} are recovered.

\section{Monte Carlo Simulations}
\label{NumericalSection}

\subsection{Local persistence probability}

%
%
\begin{figure}
\epsfxsize=85mm
\centerline{\epsffile{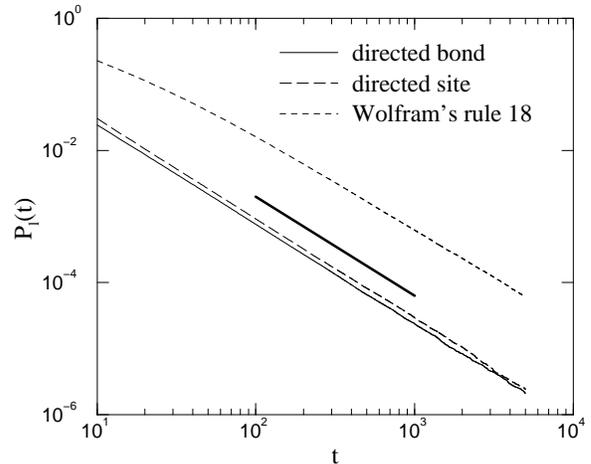}}
\caption{
Universal properties of the local persistence exponent $\theta_l$.
The probability $P_l(t)$ that a site never becomes active up to
time~$t$ is measured in a Domany Kinzel model with $1024$ sites
averaged over $10^5$ independent runs. Starting with the initial 
density $\rho(0)=0.8$ various critical points are examined. 
The results indicate a power law decay which is consistent with 
$\theta_l=1.49(2)$ in all cases. The bold line marks slope -3/2.
}  
\label{FigureUnivOne}
\end{figure}
\noindent

The local persistence probability $P_l(t)$  is defined as the
probability that in a DP process starting from random initial conditions
a given inactive site does not become active up to time~$t$. This quantity
is nontrivial because it can be seen as an infinite-point 
correlation function in the direction of time.

Notice that in contrast to Ising systems we define $P_l(t)$ as the
probability for a site not to become active rather than not to flip.
This is because there is no symmetry between active and inactive sites
in DP: Since active sites can spontaneously turn into inactive sites,
the probability for a site to remain active up to time~$t$ decays
exponentially. On the other hand $P_l(t)$, defined as the probability
for a site to remain inactive, is expected to decay more slowly since
inactive sites can only become active in presence of an active
neighboring site.

The qualitative behavior of $P_l(t)$ depends on the percolation
probability.  In the inactive phase the density of active sites decays
exponentially fast until the system enters the absorbing
state. Therefore a finite fraction of sites remain inactive 
so that $P_l(t)$ saturates at some constant value. In the active phase
an infinite percolating cluster emerges. Since the size of inactive 
islands in this cluster is finite, $P_l(t)$ will decay exponentially
in that case. At the critical point, however, there is no 
characteristic length scale, wherefore we expect $P_l(t)$ 
to decay as a power law.

%
%
\begin{figure}
\epsfxsize=85mm
\centerline{\epsffile{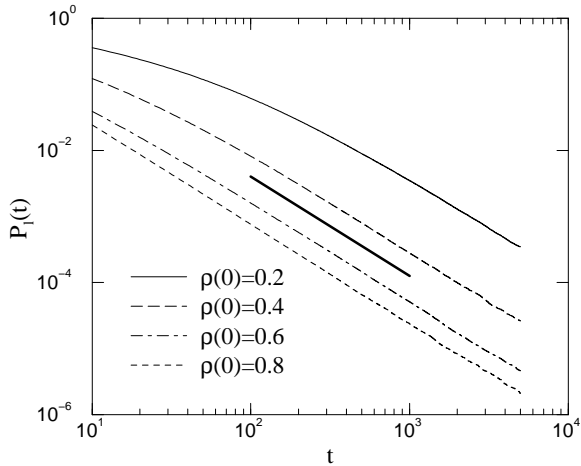}}
\caption{
Universal properties of the local persistence exponent $\theta_l$.
$P_l(t)$ is measured for directed bond percolation with various
initial densities. For $\rho(0) \geq 0.4$ an algebraic decay
with $\theta_l=1.50(2)$ is observed. In the case of very low 
initial density there seems to be a longer transient until the final
power law decay is reached. The bold line indicates the slope -3/2. 
}
\label{FigureUnivTwo}
\end{figure}
\noindent

We performed Monte Carlo simulations for various transition points and
different initial densities. In all cases a power law decay is
observed.  The local persistence exponent $\theta_l \approx 1.50(3)$
is found to be independent of the choice of the transition point as
well as the initial density (see Figs.~\ref{FigureUnivOne}
and~\ref{FigureUnivTwo}). Similar results (not reported here) are
obtained in a contact process~\cite{ContactProcess} which is a model
for DP with random sequential dynamics. Therefore our findings suggest
that $\theta_l$ is indeed a universal exponent in DP.

\begin{figure}
\epsfxsize=85mm
\centerline{\epsffile{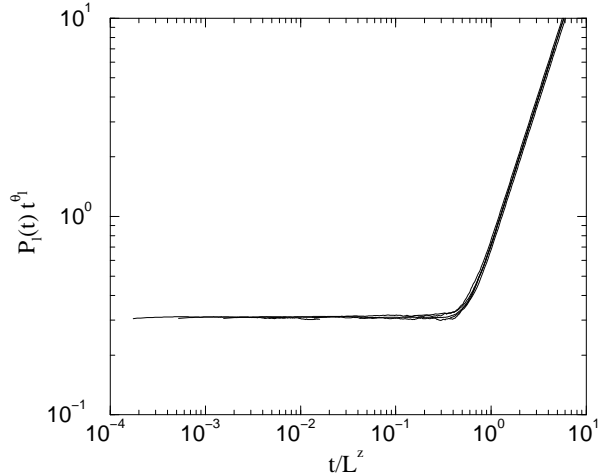}}
\caption{
Finite-size scaling function 
for the local persistence in the absorbing phase.
The scaled persistence probability $P_l(t) t^{\theta_l}$ is measured
at criticality for system sizes $L=8,16,32,\ldots,1024$ and
agains scaled time $t/L^z$. The best data collapse is 
obtained for $\theta_l=1.50(1)$.
}
\label{FigureScaling}
\end{figure}

In addition we examined the scaling properties of the local persistence
probability. According to  the usual scaling theory of
DP~\cite{DPReview}, 
we expect $P_l(t)$ to scale like
\begin{equation}
\label{ScalingFunction}
P_l(t,L,\epsilon) \sim t^{-\theta_l} \,
f(\epsilon^{\nu_{||}} t, L^{-z} t)\,,
\end{equation}
where $\epsilon=|p-p_c|$ measures the distance from the critical
point, $z=\nu_{||}/\nu_{\perp}$ is the dynamical exponent,
and $f$ is a universal scaling function with the asymptotic behavior
\begin{eqnarray}
f(0,0) &=& const \nonumber \\
f(x,0) &\sim& x^{\theta_l} \mbox{\ for \ } x\rightarrow \infty \\
f(0,y) &\sim& y^{\theta_l} \mbox{\ for \ } y\rightarrow \infty\,.
\nonumber
\end{eqnarray}
We verified the finite-size scaling of Eq.~(\ref{ScalingFunction})
at criticality. In Fig.~\ref{FigureScaling} the scaling function 
$f(0,L^{-z}t)$ is shown for various system sizes. 
The best data collapse is obtained for $\theta=1.50(1)$ 
which is our the most precise estimate for the local persistence
exponent. Attempts to relate $\theta_l$ to the DP exponents 
$\beta$, $\nu_\perp$, $\nu_{||}$ have failed wherefore we believe 
that $\theta_l$ is an independent critical exponent.

%
%
\noindent
%
%

\subsection{Global persistence probability}
\label{GlobalPersistenceSection}

The global persistence probability $P_g(t)$ is usually defined as the
probability that the global order parameter (analogous to the total
magnetization in Ising Model) does not change its sign up to
time~$t$. It has been studied in various models as, for example, in
the Ising model~\cite{MBCS} and models in the parity conserving
class~\cite{NMGO}.  In all cases it was found that $P_g(t)$ decays
algebraically with an exponent $\theta_g$ independent of $\theta_l$
and the other scaling exponents of the system.

In DP, however, the global order parameter -- the density of active
sites $\rho(t)$ -- is a non-negative quantity and therefore the above
definition is not applicable. Instead we may consider the probability
that the {\em deviation} of the order parameter from its mean value
$\Delta\rho(t) = \rho(t)-\langle \rho(t) \rangle$ does not change its
sign up to time~$t$. As in the case of local persistence it turns out
that this probability depends on the sign of $\Delta\rho(t)$ since
there is no symmetry between active and inactive sites. In fact, the
probability for the deviation to remain positive up to time~$t$ decays
exponentially fast whereas the corresponding probability for negative
deviation decays more slowly.  We will therefore define $P_g(t)$ as
the probability that $\Delta\rho(t)<0$ up to time~$t$.

Usually it is very difficult to measure $\theta_g$ by Monte Carlo
simulations, the reason being that each run produces only one data
point (the first passage time) whereas the measurement of $\theta_l$
produces a number of data points of the order of the system
size $L$.  Hence a large number of runs is required to obtain good
statistics.  However, it has been shown recently~\cite{BlockSpinMethod}
that this problem can be circumvented by
introducing spin-block persistence probabilities $P_m(t)$ which are
defined as follows: Consider blocks of $m$ sites and define a block
density $\rho_m(t)$ as the average fraction of active sites in each
block. $P_m(t)$ is defined as the probability that $\Delta\rho_m(t) =
\rho_m(t)-\langle \rho(t) \rangle$ remains negative up to
time~$t$. Obviously the spin-block persistence probability $P_m(t)$
connects the special cases of local and global persistence, namely
\begin{equation}
P_1(t)=P_l(t),
\qquad
P_\infty(t)=P_g(t)\,.
\end{equation}
It was observed in Ref.~\cite{BlockSpinMethod} that $P_m(t)$ in a
Glauber model first decays as $P_g(t) \sim t^{-\theta_g}$ and then
crosses over to the power law decay $P_m(t) \sim P_l(t) \sim
t^{-\theta_l}$, where the crossover time grows with the box
size. Since the number of such boxes is of the order $L/m$,
this method allows to measure $\theta_g$ much more accurately.

\begin{figure}
\epsfxsize=85mm
\centerline{\epsffile{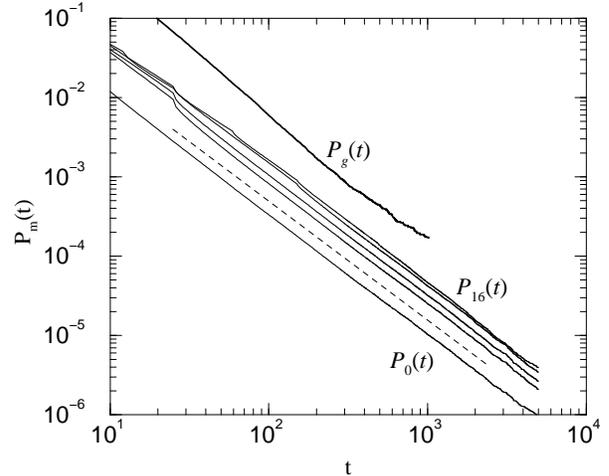}}
\caption{
Global persistence probability and Block spin method:
Decay of the persistence probability $P_m(t)$ for block sizes
$m=1,2,4,8,16$. From the slopes we obtain the estimates 
$\theta_1=1.52(2)$, $\theta_2=1.48(2)$, $\theta_4=1.49(2)$, 
$\theta_8=1.52(2)$, $\theta_{16}=1.49(2)$. The bold line
represents a direct measurement of $P_g(t)$ (shifted vertically
by a factor of $10$). The dashed line indicates slope -3/2.
}
\label{FigureBlocks}
\end{figure}

We have measured the $P_m(t)$ for block sizes $m=1,2,4,8,16$.  As can
be seen in Fig.~\ref{FigureBlocks}, $P_m(t)$ decays as $t^{-\theta_m}$
where the exponents $\theta_m \simeq 1.50(4)$ seem to be independent
of $m$, suggesting that the global and the local persistence exponents
are identical. This is surprising since in all previously known cases
$\theta_g$ was found to be smaller than $\theta_l$. For example, in
the 1d Ising model $\theta_g=1/4$ and $\theta_l=3/8$~\cite{DHP,MBCS}.

In order to support this observation, we also measured the global
persistence probability $P_g(t)$ over two decades in time
(see bold line in Fig.~\ref{FigureBlocks}).  Although this measurement
is not very accurate, there is no indication for $\theta_g$ to be
smaller than $\theta_l$ which supports the previous result. 
At present we have no explanation why the exponents $\theta_l$ 
and $\theta_g$ appear to be equal in DP.

%
%
\noindent
%
%

\subsection{Power law versus stretched exponential}
\label{StretchedSubsection}

\begin{figure}
\epsfxsize=85mm
\centerline{\epsffile{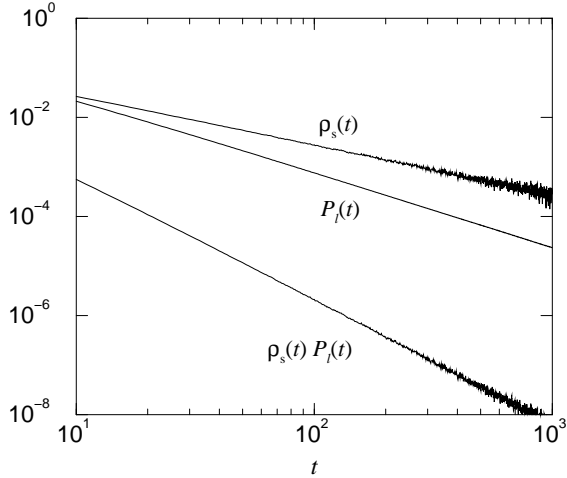}}
\caption{
Verification of the power law decay of $P_l(t)$. The probability
$\rho_s(t)$ for finding an active site near a site which was never
active until time $t$ is plotted as a function of time. It is observed
that $\rho_s(t) \sim t^{-\alpha}$ with $\alpha=1.01(2)$, indicating
that $P_l(t)$ decays indeed algebraically (see text).  }
\label{FigureNoStretch}
\end{figure}

In certain reaction diffusion models it was shown that the $P_l(t)$
decays as a stretched exponential function rather than a power
law~\cite{Cardy}.  In numerical simulations it is sometimes difficult
to distinguish between stretched exponential and power law decay. In
order to verify whether $P_l(t)$ truly decays as a power law one can
use a heuristic argument as indirect test~\cite{Cardy}: For a site
which was never active up to time~$t$ to become active in the next
time step $t+1$, it is necessary that a neighboring site is active at
time $t$. Hence $P_l(t)$ changes according to
\begin{equation}
\label{DGL}
\frac{d}{dt} P_l(t) = - P_l(t) \rho_s(t)
\end{equation}
where $\rho_s(t)$ is the probability for finding an active site near a
site which was never active until time $t$. By integrating
Eq.~(\ref{DGL}) one can easily see that $P_l(t)$ decays as a power law
only if ${\rho_s}(t) \sim t^{-1}$.  On the other hand, if
${\rho_s}(t)$ decayed as $t^{-\alpha}$ with $\alpha \neq 1$, $P_l(t)$
would decay as a stretched exponential. Thus by measuring the
exponent $\alpha$ in a Monte Carlo simulation we can verify the
observed power law decay for $P_l(t)$. Our results (see
Fig.~\ref{FigureNoStretch}) are consistent with $\rho_s(t) \sim
t^{-1}$, supporting that $P_l(t)$ actually decays algebraically.

%
%
\noindent
%
%

\section{Relation to directed percolation with an absorbing boundary}
\label{AbsorbingBoundarySection}

In this section we prove that the local persistence probability
$P_l(t)$ is exactly equal to the expectation value of a specific
observable in a DP process with an absorbing boundary. A boundary is
called absorbing if all bonds across the boundary are cut. In a $1$-d
system such an absorbing boundary can be introduced by forcing a
particular site to be inactive during the whole time evolution,
i.e. an absorbing boundary can be understood as a dry wall. The effect
of an absorbing boundary in a $1$-d DP process has been studied
recently in Refs.~\cite{EGJT96,LSMJ97,FHL97}, and it is therefore
interesting to investigate the relation between the two problems.

Let us consider a DK model with $L$ sites and periodic boundary
conditions.  Denoting by
$|\sigma\rangle=\{\sigma_1,\sigma_2,\ldots,\sigma_L\}$ basis vectors
in configuration space, the transfer matrix $\transfer$ of the DK
model is defined by
\begin{equation}
\label{TransferMatrix}
\langle \sigma^\prime(t+1)|\transfer|\sigma(t)\rangle = 
\prod_{i=1}^L \tau\Bigl(\sigma_i(t+1)|\sigma_{i-1}(t),
\sigma_{i+1}(t)\Bigr)\,.
\end{equation}
Furthermore let us define vectors for the absorbing
configuration $|0\rangle$, the state $|a\rangle$ where all sites are
active, and the sum over all configurations $|1\rangle$:
\begin{eqnarray}
|0\rangle &=& \{0,0,\ldots,0,0\} \nonumber\\
|a\rangle &=& \{1,1,\ldots,1,1\} \\
|1\rangle &=& \sum_{\sigma_1,\sigma_2,\ldots,\sigma_L}
            |\sigma \rangle \nonumber
\end{eqnarray}
Using this notation the transfer matrix obeys
\begin{equation}
\transfer|0\rangle = |0\rangle \,, \qquad
\langle 1|\transfer = \langle 1|\,,
\end{equation}
i.e. the absorbing state is a  ground state of the DK model and the
transfer matrix conserves probability. Let us now select an arbitrary
site $j$ and define local operators $\smat_0$, $\pmat_0$, and $\wmat$ by
\begin{eqnarray}
\langle \sigma^\prime |\smat_0| \sigma\rangle &=&
\delta_{\sigma^\prime_j,0} \prod_{i=1, i \neq j}^L 
\delta_{\sigma^\prime_i,\sigma_i}
\nonumber \\
\langle \sigma^\prime |\pmat_0| \sigma\rangle &=&
\delta_{\sigma^\prime_j,0} \prod_{i=1}^L
\delta_{\sigma^\prime_i,\sigma_i}
\\
\langle \sigma^\prime |\wmat| \sigma\rangle &=&
\tau(0|\sigma_{j-1},\sigma_{j+1}) 
\prod_{i=1}^L \delta_{\sigma^\prime_i,\sigma_i} \nonumber \,\,.
\end{eqnarray}
$\smat_0$ turns site $j$ into the inactive state, $\pmat_0$ projects
onto states where site $j$ is inactive, and $\wmat$ is a diagonal
weight operator to be explained below. Notice that $\smat_0$ conserves
probability whereas $\pmat_0$ does not. Using these notations we can
now express the local persistence probability~$P_l(t)$ as
\begin{equation}
\label{TransferLocalPersistence}
P_l(t) = \langle 1 | (\pmat_0 \transfer)^t |i\rangle\,.
\end{equation}
Here $|i\rangle$ denotes the initial probability distribution where
the average over different initial configurations is taken into
account.  For example, the average over random initial conditions with
particle density $\rho$ corresponds to the initial distribution
\begin{equation}
\label{RandomInitial}
|i\rangle=\sum_{\sigma_1,\sigma_2,\ldots,\sigma_L} \, \prod_{i=1}^L \,
[\rho\sigma_i+(1-\rho)(1-\sigma_i)]\,|\sigma\rangle\,.
\end{equation}
In Eq.~(\ref{TransferLocalPersistence}) the projector $\pmat_0$ 
removes all space-time histories which would turn site~$j$
into the active state giving just the local persistence 
probability $P_l(t)$. Notice that this expression is already
properly normalized.

Now consider a DP process with an absorbing boundary inserted at site
$j$.  This process is described by a different transfer matrix
$\tilde{\transfer}$ where site $j$ is forced to be inactive:
\begin{equation}
\tilde{\transfer}=\smat_0\transfer \,.
\end{equation}
One can easily show that $\tilde{\transfer}$ is related to the full
transfer matrix $\transfer$ by$ \pmat_0\transfer=\tilde{\transfer}\wmat$
which implies that 
\begin{equation}
\label{WEquation}
P_l(t) = \langle 1| (\tilde{\transfer}\wmat)^t |i\rangle\,.
\end{equation}
Therefore the local persistence probability $P_l(t)$ is {\em exactly}
equal to the expectation value of the diagonal operator~$\wmat$
measured before each update in a DP process with an absorbing
boundary. Eq.~(\ref{WEquation}) can be written as
\begin{equation}
\label{DryWallExact}
P_l(t) = \left\langle
\prod_{t^\prime=0}^{t-1}
\tau(0|\sigma_{j-1}(t^\prime),\sigma_{j+1}(t^\prime))
\right\rangle
\end{equation}
where $\langle\rangle$ denotes the average over many independent
realizations of a DP process with an absorbing boundary at site $j$
combined with an independent average over initial configurations
according to the probability distribution~$|i\rangle$. Note that
$\tau(0|\sigma_{j-1}(t^\prime),\sigma_{j+1}(t^\prime))$ refers to the
transition probability~(\ref{TransitionProbabilities})
of the DK model without dry wall at site $j$.

Because of the dry wall at site $j$, 
$\sigma_{j-1}(t)$ and $\sigma_{j+1}(t)$ are uncorrelated
in an infinite system.
Assuming that events where $\sigma_{j-1}(t)=\sigma_{j+1}(t)=1$ are rare
we may therefore approximate Eq.~(\ref{DryWallExact}) by
\begin{equation}
\label{DryWallApprox}
P_l(t) \simeq \left\langle
\exp \left[ -\gamma \sum_{t^\prime=0}^{t-1} \sigma_{j+1}(t^\prime)
\right]
\right\rangle\,,
\end{equation}
where $\gamma=-2\log(1-p_1)$. Thus $P_l(t)$ is related to the
integrated activity next to the boundary. As mentioned before,
the surface activity of a $1$-d DP process with an absorbing boundary 
has been carefully analyzed by series expansions and MC simulations 
in Refs.~\cite{EGJT96,LSMJ97}. It was observed that the activity 
next to the boundary decays according to a power law
\begin{equation}
\left\langle  \sigma_{j \pm 1}(t) \right\rangle \sim
t^{-\beta^\prime/\nu_{||}}\,.
\end{equation}
Furthermore, it has been conjectured that 
$\beta^\prime=\nu_{||}-1$ which means that
\begin{equation}
\label{DryWallPowerLaw}
\left\langle \sum_{t^\prime=0}^{t-1} \sigma_{j+1}(t^\prime) 
\right\rangle \sim t^{1/\nu_{||}}\,.
\end{equation}
If the average $\langle \rangle$ in Eq.~(\ref{DryWallApprox}) 
commuted with the exponential function, Eq.~(\ref{DryWallPowerLaw})
would imply that $P_l(t)$ decays asymptotically 
as a {\em stretched exponential}
\begin{equation}
\label{StretchedExponential}
P_l(t) \sim \exp(-\gamma \, t^{1/\nu_{||}})
\end{equation}
rather than a power law. 
However, our numerical results strongly suggest that $P_l(t)$
does indeed decay as a power law (see Sect.~\ref{StretchedSubsection}). 
In fact, a stretched exponential of 
the form~(\ref{StretchedExponential}) is in obvious contradiction with the 
simulation data. Thus the average operation $\langle\rangle$ 
certainly not commutes with the exponential function. Nevertheless
we are left with a puzzle: For both Eqs.~(\ref{DryWallApprox}) 
and~(\ref{DryWallPowerLaw}) to decay algebraically, a delicate
mechanism in the exponential function is needed, i.e. higher cumulants 
of the integrated surface activity have to match in a very specific way.

\section{Relation to a directed percolation process with an active
source}
\label{ActiveSourceSection}
 
%
%
\begin{figure}
\epsfxsize=90mm
\centerline{\epsffile{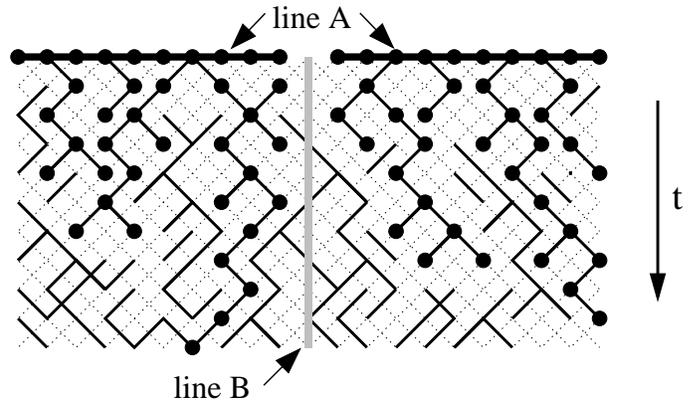}}
\vspace{3mm}
\caption{
Mapping of the local persistence problem onto a directed bond percolation
process with a steady source in the center. The figure shows a
particular realization of open bonds (solid lines) on a tilted
square lattice (dotted lines) in directed bond percolation. 
Active sites are marked by bold dots. Initially all sites are
in the active state (line A). Activity then percolates through the
system without touching site $j$ (line B). In reverse
time direction this means that there is no open path backwards from
line B to line A.
}
\label{FigureLineAB}
\end{figure}
\noindent

We will now show that the local persistence probability $P_l(t)$ can
be related to a return probability in a 1d DP process with a pointlike
active source. More precisely we will show that $P_l(t)$ is equal to
the probability for a DP process with an active source to return to a
state where all sites except for the source are inactive.  We will
prove this relation exactly in the case of directed bond percolation.

The mapping can qualitatively be understood  as follows: Consider a
particular realization of a DP process (a single MC run) starting from
initial conditions where all sites except for site $j$ are active.
Let us assume that in this particular realization
site~$j$ has never become active up to time $t_1$. Such a realization is
shown schematically in Fig.~\ref{FigureLineAB}. Obviously there
are no open paths from the horizontal line $t=0$ (line A in
Fig.~\ref{FigureLineAB}) to the vertical 
line at site $j$ given by $0 < \tilde{t} \leq t_1$ (line B).
Conversely, there is no open path {\em backwards} in time from line B
to line A. In the special case of directed bond percolation we may now
consider a DP process in reverse time direction using the same
realization of open and closed bonds. Furthermore, let us assume that
we force site~$j$ to be active along line B, i.e. we impose an active
source at this location. If there is no open path from B to A, 
activity will not percolate from line B to line~A in the
`time-reversed' process. In other words, the reverse
process returns to its initial condition where all sites except for
the source are inactive.

We now prove this mapping for the special case of directed 
bond percolation. More precisely we show that in this case
the probability $R(t)$ of a DP process with an active source to
return to its initial condition where all sites except for site~$j$ are
inactive is {\em exactly} equal to the persistence probability $P_l(t)$.
Using the notations of Sect.~\ref{AbsorbingBoundarySection}, let us
introduce two further operators $\smat_1$ and~$\umat$:
\begin{eqnarray}
\langle \sigma^\prime |\smat_1| \sigma\rangle &=&
\delta_{\sigma^\prime_j,1} \prod_{i=1, i \neq j}^L 
\delta_{\sigma^\prime_i,\sigma_i}
\nonumber \\
\langle \sigma^\prime |\umat| \sigma\rangle &=&
 \prod_{i=1}^L (1-\delta_{\sigma^\prime_i,1}\delta_{\sigma_i,1})  \,.
\end{eqnarray}
The operator $\smat_1$ turns site~$j$ into the active state whereas
the operator~$\umat$ is a symmetric transformation matrix. One can easily
verify that for $p_2=p_1(2-p_1)$, i.e., for directed bond percolation,
the following relation holds:
\begin{equation}
\label{TCommute}
\umat\transfer=\transfer^T\umat\,.
\end{equation}
Furthermore we have
\begin{equation}
\label{SPCommute}
\umat\pmat_0=\smat_1^T\umat \,, \qquad  \umat\smat_0=\smat_1^T\umat \,.
\end{equation}
By commuting the matrix $\umat$ to the right and transposing the
resulting
expression, we can now rewrite the local persistence probability 
$P_l(t)$ in Eq.~(\ref{WEquation}) as
\begin{eqnarray}
P_l(t)  &=& \langle 1| (\pmat_0\transfer)^t |i\rangle \nonumber \\
        &=& \langle 0| \umat (\pmat_0\transfer)^t \smat_0 |a \rangle
\nonumber \\
        &=& \langle 0| (\smat_1^T \transfer^T)^t \smat_0^T \umat
|a\rangle \\
        &=& \langle a| \umat \smat_0 (\transfer \smat_1)^t |0\rangle
\nonumber \\
        &=& \langle 0| \smat_0 (\transfer \smat_1)^t |0 \rangle =
R(t)\,, \nonumber
\end{eqnarray}
where we assumed the initial condition $|i\rangle = \smat_0|a\rangle$
for the persistence measurement in which all sites except for 
site~$j$ are active. We also used the relations
$\langle 1| =\langle 0| \umat $ and 
$ \langle a| \umat = \langle 0|$. The resulting expression
$\langle 0| \smat_0 (\transfer \smat_1)^t |0 \rangle = R(t)$ is precisely
the return probability of a DP process to its initial condition
with an active source at site $j$, which completes the proof. 
Notice that similar arguments were used in Ref.~\cite{DHP} in order
to derive the persistence exponent in the Glauber model.

Although this proof holds only for the case of directed bond
percolation,
the relation seems to be more general. In fact, we verified numerically
that $P_l(t)$ and $R(t)$ exhibit the same power law behavior for
various transition points in the DK model.

\section{Conclusions}
\label{SummarySection}
In the present work the problem of local and global persistence in
directed percolation has been studied numerically. The results suggest
that both the local and global persistence probabilities exhibit an
algebraic decay in time at the critical point.  The corresponding
critical exponents $\theta_l$ and $\theta_g$ are universal and
independent of the initial density. The universality of these
exponents is stronger than in the Ising model where an explicit
dependence on the initial density is found for $\theta_l$. The reason
might be an exponentially fast decaying memory of initial conditions
in DP since active sites can spontaneously become inactive.
Surprisingly, both $\theta_l$ and $\theta_g$ seem to be equal within
numerical errors which contrasts to previously known cases where
it was found that $\theta_g<\theta_l$. We carefully analyzed our data
in order to rule out any stretched exponential decay of the persistence
probability.

In order to consider the problem in a more general context, we related
the local persistence probability to certain observables in a DP
process with a dry and a wet wall, respectively. Introducing a
transfer matrix formalism these relations were proven exactly
for particular realizations of DP.

Various questions remain open. First of all, we do not know whether
$\theta_l$ and $\theta_g$ are in fact independent from the usual
critical exponents in DP. The numerical value $\theta_l
\approx 1.50(1)$ suggests that the exact value could be $3/2$,
a possibility which cannot be ruled out as an integer exponent was
also observed in a DP process with an absorbing
boundary~\cite{EGJT96,LSMJ97}. Furthermore, it is not clear why
$\theta_l$ and $\theta_g$ appear to be identical in DP.
Finally, it would be interesting to investigate 
the same problem in higher dimensions. \\[1mm]
Acknowledgements: We thank P.~Grassberger, K.~B.~Lauritsen,
D.~Mukamel and C.~Sire for interesting discussions. 
H. M. K. thanks the Max-Planck-Institut
f\"ur Physik komplexer Systeme, Dresden for hospitality where parts of
this work were done.

\vspace{-5mm}


\end{document}